\documentclass[showpacs,amssymb,aps,twocolumn]{revtex4}
\usepackage{amsmath}
\usepackage{graphicx}
\usepackage[latin1]{inputenc}
\begin{document}

\title{Thermal Operator Representation of Finite Temperature Graphs II}

\author{F. T. Brandt$^{a}$, Ashok Das$^{b}$, Olivier Espinosa$^{c}$,
  J. Frenkel$^{a}$ and Silvana Perez$^{d}$}
\affiliation{$^{a}$ Instituto de Física, Universidade de São
Paulo, São Paulo, BRAZIL}
\affiliation{$^{b}$ Department of Physics and Astronomy,
University of Rochester,
Rochester, New York 14627-0171, USA}
\affiliation{$^{c}$ Departamento de Física, Universidad
Técnica Federico Santa María, Casilla 110-V, Valparaíso, CHILE}
\affiliation{$^{d}$ Departamento de Física, 
Universidade Federal do Pará, 
Belém, Pará 66075-110, BRAZIL}

\bigskip

\begin{abstract}

Using the mixed space representation, we extend our earlier analysis
to the case of Dirac and gauge fields and show that in the absence of
a chemical potential, the finite temperature Feynman diagrams can be
related  to the
corresponding zero temperature graphs through a thermal operator. At
non-zero chemical potential we show explicitly in the case of the
fermion self-energy that such a factorization is violated because of
the presence of a singular contact term. Such a temperature dependent
term which arises only at finite density and has a quadratic mass
singularity cannot be related, through a regular thermal operator, to
the fermion self-energy at zero temperature which is infrared finite.
Furthermore, we show that the thermal radiative
corrections at finite density have a screening effect for the chemical
potential leading to a finite renormalization of the potential.

\end{abstract}


\maketitle

\section{Introduction}

In an earlier paper \cite{silvana} (referred to as {\bf I}), we gave a
simple derivation of
an interesting relation \cite{Espinosa:2003af,
  Espinosa:2005gq,Blaizot:2004bg} between finite temperature Feynman
graphs and 
the corresponding zero temperature graphs  within the context of a
scalar field theory in real as well as in imaginary time
formalisms. We showed that the derivation is particularly
simple if one uses a mixed space representation of the graphs in the $(t,
\vec{p}\ )$ space \cite{das:book97,Bedaque:1993fa,Das:1997gg} and the
proof of the correspondence is particularly
simple in the closed time path formalism
\cite{das:book97,Schwinger:1961qe,Bakshi:1962dvKeldysh:1964ud}.
Explicitly, for any
$N$-point graph (at any loop) in a scalar field theory at finite
temperature, the relation can be written as
\begin{eqnarray}
& &\Gamma_{N}^{(T)} = \int \prod_{i=1}^{I}
  \frac{d^{3}k_{i}}{(2\pi)^{3}} \prod_{v=1}^{V} 
  (2\pi)^{3} \delta_{v}^{(3)} (k,p) \gamma_{N}^{(T)}\nonumber\\
& & = \int \prod_{i=1}^{I} \frac{d^{3}k_{i}}{(2\pi)^{3}} \prod_{v=1}^{V}
  (2\pi)^{3} \delta_{v}^{(3)} (k,p) {\cal O}^{(T)}
  \gamma_{N}^{(0)},\label{tor}
\end{eqnarray}
where 
\begin{equation}
{\cal O}^{(T)} = \prod_{i=1}^{I} \left(1 + n_{i} (1 -
S_{i})\right),\label{toB}
\end{equation}
with $E_{i} = \sqrt{\vec{k}^{2} + m^{2}},\; n_{i} = n (E_{i})$
denoting the Bose-Einstein distribution function associated with the
internal propagators and
$S_{i} = S (E_{i})$ represents a reflection operator that changes
$E_{i}\rightarrow - E_{i}$. In (\ref{tor}), $I$ characterizes the
number of internal propagators, $V$ the total number of vertices in
the graph (with the usual relation for the number of loops $L= I - V
+1$) and $\delta_{v}^{(3)} (k,p)$ enforces the conservation of
momentum at the vertex $v$. We denote the internal and the external
three momenta of a graph generically by $\vec{k}, \vec{p}$
respectively. Furthermore, $\gamma_{N}^{(T)}$ represents the integrand
of the finite temperature graph (after the internal time coordinates
have been integrated in the mixed space) so that it has the dependence
\begin{equation}
\gamma_{N}^{(T)} = \gamma_{N}^{(T)} (T, \vec{k}_{i}, t_{\alpha}),
\end{equation}
where $t_{\alpha}, \alpha = 1,2,\ldots , N$ denote the external
time coordinates of the graph while $\gamma_{N}^{(0)}$ is the
integrand of the same graph at zero temperature (with the internal
time coordinates integrated). The operator (\ref{toB}) relating the
integrands of the two graphs was termed the {\em thermal operator} and
the most important property of this operator is that it is independent
of time coordinates and carries the entire temperature dependence of
the (finite temperature) graph. This interesting result is
calculationally quite useful and allows us to study directly many
questions of interest at finite temperature such as Ward identities
and 
analyticity \cite{Weldon:1982aq,braaten,frenkel}. In {\bf I}, we had
shown  that this simple
relation arises as a consequence of the factorization of the finite
temperature propagator in the scalar field theory into a basic thermal
operator acting on the zero temperature propagator and had studied various
properties associated with this thermal operator. In particular, we
had shown that the basic thermal operator for the propagator
corresponds to a projection operator that projects onto the space of
periodic functions. (We recall that while the finite temperature propagator
for the scalar field satisfies periodic conditions following from the
Kubo-Martin-Schwinger condition \cite{Kubo:1957xxMartin:1959jp}, the
zero temperature propagator  does not.) For a
complex scalar field with a chemical potential, on the other hand, we
showed that the basic thermal operator is much more complex involving
time derivative terms. In this case, we could not give a general
proof of a thermal operator representation such as (\ref{tor})
although we showed, for specific complicated graphs, that a nontrivial
factorization nonetheless arises.

In this paper, we extend our analysis in {\bf I} to theories involving
fermions as well as gauge theories. The analysis for gauge theories is
particularly of interest since the interaction terms (non-Abelian
three point interaction as well as the interaction of the ghost
fields) involve derivative terms. We find in all cases that if there
is no chemical potential present, a thermal operator representation
for finite temperature graphs naturally follows. On the other hand,
for a fermion with a chemical potential, as in the case of the complex
scalar field discussed in {\bf I}, the basic thermal operator is complicated
involving time derivatives and we find that a thermal operator
representation for graphs fails. This failure is traced to the fact
that in such theories, the self-energy develops a quadratic mass
singularity because of radiative corrections at finite density. The
paper is organized as
follows. In section {\bf II}, we discuss fermion theories at finite
temperature (without a chemical potential) and show that the basic
factorization of the thermal propagator arises much as in the scalar
field theory. The proof of the thermal operator representation for an
interacting theory involving scalar and fermion fields is direct in
the closed time path formalism which we discuss. In section {\bf III},
this analysis is extended to gauge theories where we show that the
basic factorization of the thermal propagator leads to a thermal
operator representation for any graph at finite temperature in spite
of interaction terms involving derivatives. The thermal operator
representation is explicitly worked out for the contribution of the
ghost loop to the self-energy of the gauge field. In section {\bf IV},
we study an interacting theory of gauge fields and fermions with a
chemical potential (for example, QED at finite density) and show that
in this case the basic factorization of the thermal propagator for the
fermion involves a dependence on time derivatives. In this case, the
basic thermal operator can also be written equivalently as one without
a time derivative but with a matrix structure. We work
out the fermion self-energy in this theory explicitly and show that a
thermal 
operator representation fails. We trace this failure in section {\bf
  V} to the fact that the quantum corrections in this theory lead to a
quadratic mass singularity at finite density. By analyzing the pole of
the fermion propagator in this theory, we show that the chemical
potential has a finite renormalization due to radiative corrections
and we discuss some interesting aspects of this phenomenon (see, for
example,  ref. \cite{bijlsma} for a
discussion from the point of view of the renormalization group
evolution). We conclude with a brief
summary in section {\bf VI}. In
appendix {\bf A}, we study the  $0+1$ dimensional Chern-Simons QED to
 bring out some interesting features of the thermal operator
 representation in lower dimensions while appendix {\bf B} describes
 briefly the derivation of some of the formulae used in the text.

\section{Fermions (without chemical potential)}

In this section, we will study an interacting theory of scalar and
fermion fields at finite temperature (without chemical potential). As
we have shown in {\bf I}, the proof of the thermal operator representation
for any graph is direct in the closed time path formalism
\cite{das:book97,Schwinger:1961qe,Bakshi:1962dvKeldysh:1964ud} . 
Therefore, for simplicity, we will discuss the theory in this formalism although
everything we say also holds in the imaginary time formalism. Indeed,
in reference \cite{Espinosa:2005gq} the validity of the thermal
operator representation has been shown to hold when there is no
chemical potential.
Let us consider the theory described by the Lagrangian density
\begin{widetext}
\begin{equation}
{\cal L} =  \bar{\psi}\left(i \partial\!\!\!\slash - m\right) \psi +
\frac{1}{2} \partial_{\mu}\phi \partial^{\mu}\phi - \frac{M^{2}}{2}
\phi^{2} - g \bar{\psi}\psi \phi - \frac{\lambda}{4!} \phi^{4}.
\end{equation}
\end{widetext} 
The factorization of the scalar propagator has already been discussed
in {\bf I} and we simply recapitulate here the essential results. In the
closed time path formalism, the propagator has a $2\times 2$ structure
which can be written as
\begin{equation}
\Delta^{(T)} (t,E) = {\cal O}_{\rm B}^{(T)} \Delta^{(0)} (t,E),
\end{equation}
where the $2\times 2$ matrix structure of the propagator is labelled
at any temperature as
\begin{equation}
\Delta^{(T)}(t,E) = \left(\begin{array}{cc}
\Delta_{++}^{(T)} (t, E) & \Delta_{+-}^{(T)} (t,E)\\
\noalign{\vskip 2pt}
\Delta_{-+}^{(T)} (t,E) & \Delta_{--}^{(T)} (t,E)
\end{array}\right),\label{scalarfactorization}
\end{equation}
and the basic thermal operator is the scalar operator
\begin{equation}
{\cal O}_{\rm B}^{(T)} (E) = 1 + n_{\rm B} (E) \left(1 - S (E)\right), \label{tob}
\end{equation}
where 
\begin{equation}
n_{\rm B} (E) = \frac{1}{e^{\frac{E}{T}} - 1},\quad E = E(M) =
\sqrt{\vec{p}^{\ 2} + M^{2}}.
\end{equation}
The components of the propagator at zero temperature have the
following explicit forms in the mixed space
\begin{eqnarray}
& &\!\!\!\!\!\Delta_{++}^{(0)} (t, E) = L(\epsilon)\,
\frac{1}{2E}\left[\theta (t) e^{-i(E-i\epsilon)t} + \theta (-t)
  e^{i(E-i\epsilon)t}\right],\nonumber\\
& &\!\!\!\!\!\Delta_{+-}^{(0)} (t,E) = \frac{1}{2E} e^{iEt},\quad
\Delta_{-+}^{(0)} (t,E) = \frac{1}{2E}
e^{-iEt},\label{scalarcomponents}\\ 
& &\!\!\!\!\!\Delta_{--}^{(0)} (t,E) = L(\epsilon)\,
\frac{1}{2E}\left[\theta (t) e^{i(E+i\epsilon)t} + \theta (-t)
  e^{-i(E+i\epsilon)t}\right],\nonumber
\end{eqnarray} 
where the operator $L(\epsilon)$ takes the limit $\epsilon\rightarrow 0$.

For fermions, on the other hand, we know that the components of the
$2\times 2$ matrix propagator (in the closed time path formalism) at
finite temperature have the momentum space representation
\begin{equation}
S^{(T)} (p) = \left(\begin{array}{cc}
S_{++}^{(T)} (p) & S_{+-}^{(T)} (p)\\
\noalign{\vskip 2pt}
S_{-+}^{(T)} (p) & S_{--}^{(T)} (p)
\end{array}\right),
\end{equation}
with ($L(\epsilon)$ is the operator taking the limit
$\epsilon\rightarrow 0$ introduced earlier)
\begin{eqnarray}
& &S_{++}^{(T)} (p) =  (p\!\!\!\slash + m)\Big(L(\epsilon)\,
\frac{i}{p^{2} - m^{2} + i\epsilon}\nonumber\\
& &\qquad\qquad\qquad\qquad\qquad - 2\pi n_{\rm F} (|p_{0}|)
  \delta (p^{2}-m^{2})\Big),\nonumber\\
& &S_{+-}^{(T)} (p) =  2\pi (p\!\!\!\slash + m)\left(\theta (-p_{0}) -
  n_{\rm F} (|p_{0}|)\right)\delta (p^{2}-m^{2}),\nonumber\\
& &S_{-+}^{(T)} (p) =  2\pi (p\!\!\!\slash + m)\left(\theta (p_{0}) -
  n_{\rm F} (|p_{0}|)\right)\delta (p^{2}-m^{2}),\nonumber\\
& &S_{--}^{(T)} (p) =  (p\!\!\!\slash +
  m)\Big(-L(\epsilon)\,\frac{i}{p^{2}-m^{2}-i\epsilon}\nonumber\\
& &\qquad\qquad\qquad\qquad\qquad - 2\pi n_{\rm F} (|p_{0}|)
  \delta (p^{2}-m^{2})\Big).\label{S}
\end{eqnarray}
Here $n_{\rm F} (|p_{0}|)$ represents the Fermi-Dirac distribution
function
\begin{equation}
n_{\rm F} (|p_{0}|) = \frac{1}{e^{\frac{|p_{0}|}{T}} + 1},
\end{equation}
and the temperature dependent terms reflect the anti-periodicity
condition satisfied by the fermion propagator.

The components of the fermion propagator in (\ref{S}) can be Fourier
transformed in the energy variable to give
\begin{eqnarray}
S^{(T)} (t,\vec{p}\ ) & = & \int_{-\infty}^{\infty}
\frac{dp_{0}}{2\pi}\ e^{-ip_{0}t}\ S^{(T)} (p)\nonumber\\
 & = & {\cal O}_{\rm F}^{(T)} (E) S^{(0)} (t,\vec{p}),
\end{eqnarray}
where $E = E(m)= \sqrt{\vec{p}^{\ 2} + m^{2}}$ and 
\begin{equation}
{\cal O}_{\rm F}^{(T)} (E) = 1 - n_{\rm F} (E) (1 - S (E)) .\label{tof}
\end{equation}
The components of the zero temperature propagator have the explicit
forms
\begin{eqnarray}
& & S_{++}^{(0)} (t,\vec{p}\ ) = L(\epsilon)\,
  \frac{1}{2E}\left[\theta(t)
  A(E)e^{-i(E-i\epsilon)t}\right.\nonumber\\
& &\qquad\qquad\qquad\qquad\left. + \theta (-t)
  B(E) e^{i(E-i\epsilon)t}\right],\nonumber\\
& &S_{+-}^{(0)} (t,\vec{p}\ ) = \frac{1}{2E} B(E) e^{iEt}, 
  S_{-+}^{(0)} (t,\vec{p}\ ) = \frac{1}{2E} A(E) e^{-iEt},\nonumber\\
& &S_{--}^{(0)} (t,\vec{p}\ ) = L(\epsilon)\,
  \frac{1}{2E}\left[\theta (t) B (E)
  e^{i(E+i\epsilon)t}\right.\nonumber\\
& &\qquad\qquad\qquad\qquad\left. + \theta (-t)
  A (E) e^{-i(E+i\epsilon)t}\right],\label{fermioncomponents}
\end{eqnarray}
where
\begin{equation}
A (E) = \gamma^{0}E - \vec{\gamma}\cdot \vec{p} + m,\quad B (E) =
-\gamma^{0}E - \vec{\gamma}\cdot \vec{p} + m.\label{ab}
\end{equation}
It is worth remarking here that, as in the case of the scalar
propagator, it is easy to verify that the basic thermal operator in
(\ref{tof}) is a projection operator, namely,
\begin{equation}
\left({\cal O}_{\rm F}^{(T)} (E)\right)^{2} = {\cal O}_{\rm F}^{(T)}
(E),\label{projection}
\end{equation}
and in the present case projects onto functions satisfying
anti-periodicity properties.

\begin{figure}[h!]
\includegraphics[scale=0.4]{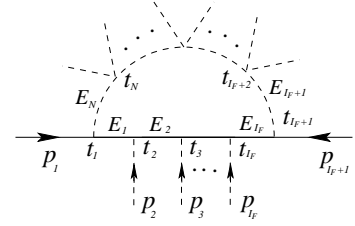}
\caption{A typical one-loop diagram involving fermions (solid lines) and scalar
  fields (dashed lines).
For simplicity, the vertices are all assumed to be of ``$+$'' type.}
\label{f1}
\end{figure}

Thus, we see that in spite of a matrix structure (from the Dirac gamma
matrices) of the fermion propagator, the thermal propagator factorizes
in terms of a basic thermal operator (\ref{tof}) which is a scalar
quantity much like in the case of the scalar field
theory. Furthermore, it is independent of the time coordinates and as
a result, the thermal operator representation for any graph can be
obtained as follows. First, let us suppose that we have a graph with only
external vertices (that is, a one loop graph). A typical $N$-vertices graph
involving fermion propagators will have the general form shown in
Fig. \ref{f1}. (The external vertices can be of ``$\pm$'' type, but we
choose all 
of them to be of ``$+$'' for illustrative purposes only. The same
derivation will go through for vertices of any type since the basic
thermal operators in (\ref{tob}) and (\ref{tof}) are scalar quantities
and have the same form for any component of the propagator.)
\noindent In this case, at finite temperature, the value of the graph
can be written as (we set $g=1=\lambda$ and ignore all the multiplicative
factors coming from the vertices for simplicity. 
The external momenta are all assumed to be flowing
into a vertex, and the internal momentum $k_i$ flows from vertex $i$ 
to vertex $i+1$ and we identify $t_{N+1} = t_{1}, k_{N+1}=k_{1}, p_{N+1}=p_{1}$.)
\begin{equation}
\Gamma_{N}^{(T)} = \int \prod_{i=1}^{N} \frac{d^{3}k_{i}}{(2\pi)^{3}}
  (2\pi)^{3} \delta^{(3)} (k_{i}-k_{i+1} + p_{i+1})
  \gamma_{N}^{(T)},
\end{equation}
with
\begin{eqnarray}
\gamma_{N}^{(T)} & = & \prod_{i=1}^{I_{\rm F}} S_{++}^{(T)}
  (t_{i}-t_{i+1},k_{i}) \prod_{i=I_{\rm F}+1}^{N} \Delta_{++}^{(T)}
  (t_{i}-t_{i+1}, E_{i})\nonumber\\
 & = & \prod_{i=1}^{I_{\rm F}} {\cal O}_{\rm F}^{(T)} (E_{i})
  S_{++}^{(0)} (t_{i}-t_{i+1},k_{i})\nonumber\\
&  & \qquad\times \prod_{i=I_{\rm F}+1}^{N}
  {\cal O}_{\rm B}^{(T)} (E_{i}) \Delta_{++}^{(0)} (t_{i}-t_{i+1},
  E_{i})\nonumber\\
& = & {\cal O}^{(T)} \prod_{i=1}^{I_{\rm F}} S_{++}^{(0)}
  (t_{i}-t_{i+1}, k_{i}) \\ 
&  & \qquad\times \prod_{i=I_{\rm F}+1}^{N}
  \Delta_{++}^{(0)} (t_{i}-t_{i+1},E_i)\nonumber\\
& = & {\cal O}^{(T)} \gamma_{N}^{(0)},
\end{eqnarray}
where we have identified the thermal operator for the graph as
\begin{equation}
{\cal O}^{(T)} = \prod_{i=1}^{I_{\rm F}} {\cal O}_{\rm F}^{(T)}
(E_{i})  \prod_{i=I_{\rm F}+1}^{N} {\cal O}_{\rm B}^{(T)}
(E_{i}).\label{toF}
\end{equation}
The basic thermal operators ${\cal O}_{B}^{(T)} (E), {\cal O}_{\rm
  F}^{(T)} (E)$ are defined in Eqs. (\ref{tob}) and (\ref{tof})
  respectively and we have to remember that in (\ref{toF})
\begin{equation}
E_{i} = \left\{\begin{array}{cl}
\sqrt{\vec{k}_{i}^{\ 2} + m^{2}} & {\rm for}\ i=1,2,\ldots , I_{\rm
  F}\\
\noalign{\vskip 2pt}
\sqrt{\vec{k}_{i}^{\ 2} + M^{2}} & {\rm for}\ i= I_{\rm F}+1,\ldots , N
\end{array}\right.
\end{equation}
As a result, we can write
\begin{equation}
\Gamma_{N}^{(T)} = \int \prod_{i=1}^{N} \frac{d^{3}k_{i}}{(2\pi)^{3}}
  (2\pi)^{3} \delta^{(3)} (k_{i}-k_{i+1} + p_{i+1})
  {\cal O}^{(T)}\gamma_{N}^{(0)},
\end{equation}
showing that in this case, the finite temperature graph can be given a
thermal operator representation. Furthermore, since in the closed time
path formalism the range of time integration at finite temperature
continues to be the same as at zero temperature and since the basic
thermal operators (\ref{tob}) and (\ref{tof}) are independent of time
coordinates (so that they can be taken outside the integral), such a
factorization of any graph with internal time coordinate (that need to
be integrated over) continues to hold and in general, for any
$N$-point graph (at any loop) at finite temperature, we have the
thermal operator representation 
\begin{equation}
\Gamma_{N}^{(T)} = \int \prod_{i=1}^{I} \frac{d^{3}k_{i}}{(2\pi)^{3}}
  \prod_{v=1}^{V}(2\pi)^{3} \delta_{v}^{(3)} (k, p)
  {\cal O}^{(T)}\gamma_{N}^{(0)},
\end{equation}
where the thermal operator follows from (\ref{toF}) to be 
\begin{equation}
{\cal O}^{(T)} = \prod_{i=1}^{I_{\rm F}} {\cal O}_{\rm F}^{(T)}
(E_{i})\prod_{i=I_{\rm F}+1}^{I} {\cal O}_{\rm B}^{(T)} (E_{i}),
\end{equation}
with $I_{\rm F}, I$ representing respectively the number of internal
fermion propagators and the total number of internal propagators.

\section{Gauge theories}

We have seen thus far that the thermal operator representation for any
Feynman graph at finite temperature holds for theories involving
scalar and fermion fields (without a chemical potential). However,
physically gauge theories are more interesting and in this section
we will discuss a non-Abelian gauge theory at finite temperature. Let
us consider a Yang-Mills theory (where the gauge fields belong to
$SU(n)$) in the Feynman gauge described by the Lagrangian density
\begin{equation}
{\cal L} = - \frac{1}{4} F_{\mu\nu}^{a} F^{\mu\nu, a} - \frac{1}{2}
\left(\partial\cdot A^{a}\right)^{2} + \partial^{\mu}\bar{c}^{a}
D_{\mu}c^{a},\label{gaugelagrangian}
\end{equation}
where $a=1,2,\ldots , n^{2}-1$ and (we set the coupling to unity for
simplicity) 
\begin{eqnarray}
D_{\mu}c^{a} & = & \partial_{\mu}c^{a} + f^{abc} A_{\mu}^{b}
c^{c},\nonumber\\
F_{\mu\nu}^{a} & = & \partial_{\mu}A_{\nu}^{a} -
\partial_{\nu}A_{\mu}^{a} + f^{abc} A_{\mu}^{b}A_{\nu}^{c}.
\end{eqnarray}
In this case, in the closed time path formalism, the gauge and the
ghost propagators at finite temperature have the momentum space
representation
\begin{eqnarray}
D_{\mu\nu;\alpha\beta}^{ab (T)} (p) & = & - \eta_{\mu\nu}\delta^{ab}
\Delta_{\alpha\beta}^{(T)} (p),\nonumber\\
D_{\alpha\beta}^{ab (T)} (p) & = & \delta^{ab}
\Delta_{\alpha\beta}^{(T)} (p),\quad \alpha,\beta =
\pm,\label{gaugepropagator}
\end{eqnarray}
where $\Delta_{\alpha\beta}^{(T)} (p)$ represent the components of a
massless scalar propagator at finite temperature and have the explicit forms 
\begin{eqnarray}
\Delta_{++}^{(T)} (p) & = & 
\left(L(\epsilon)\,
\frac{i}{p^{2}+i\epsilon} + 2\pi n_{\rm B} (|p_{0}|)
\delta (p^{2})\right),\nonumber\\
\Delta_{+-}^{(T)} (p) & = & 2\pi \left(\theta (-p_{0}) + n_{\rm B}
(|p_{0}|)\right) \delta (p^{2}),\nonumber\\
\Delta_{-+}^{(T)} (p) & = & 2\pi \left(\theta (p_{0}) + n_{\rm B}
(|p_{0}|)\right) \delta (p^{2}),\label{gaugecomponents}\\
\Delta_{--}^{(T)} (p) & = & \left(-L(\epsilon)\,
\frac{i}{p^{2} -
  i\epsilon} + 2\pi n_{\rm B} (|p_{0}|) \delta
(p^{2})\right),\nonumber 
\end{eqnarray}
where $L (\epsilon)$ is the limiting operator introduced earlier.

By taking the Fourier transform of (\ref{gaugecomponents}) with
respect to $p_{0}$, we can obtain the components of the gauge and the
ghost propagators in the mixed space. We already know from
(\ref{scalarfactorization}) and (\ref{scalarcomponents}) that at
finite temperature the components of the scalar propagator factorize
in the mixed space representation. It follows, therefore, that the
components of the gauge and the ghost propagators also factorize in
the mixed space representation as 
\begin{eqnarray}
D_{\mu\nu;\alpha\beta}^{ab (T)} (t,\vec{p}\ ) & = & {\cal O}_{\rm
  B}^{(T)} (E) D_{\mu\nu;\alpha\beta}^{(0)} (t,\vec{p}\ ),\nonumber\\
D_{\alpha\beta}^{ab (T)} (t,\vec{p}\ ) & = & {\cal O}_{\rm B}^{(T)}
  (E) D_{\alpha\beta}^{(0)} (t,\vec{p}\ ),\label{gaugefactorization}
\end{eqnarray}
where the same basic thermal operator ${\cal O}_{\rm B}^{(T)} (E)$
leading to a factorization of the gauge and the ghost propagators 
coincides with that for a scalar propagator defined in (\ref{tob})
(with $E=|\vec{p}\ |$ for a massless field). Furthermore, all the
components of the propagator factorize in the same manner and the
basic thermal operator is independent of the time coordinate.
It is worth remarking at this point that we have chosen to work in 
the Feynman gauge for simplicity. In any other covariant gauge fixing,
only the Lorentz structure of the gauge propagator generalizes, but
the basic factorization of the thermal propagator continues to
hold.  Furthermore, we consider the case of a vanishing chemical potential here for simplicity.

\begin{figure}[h!]
\includegraphics[scale=0.5]{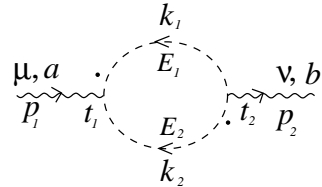}
\caption{The ghost loop contribution to the self-energy of the gauge
  boson. The ``dots'' represent the action of the time derivatives.}
\label{f2}
\end{figure}

Since the gauge and the ghost propagators factorize in the same way as
in a scalar field theory, the thermal operator representation of any
graph at finite temperature would seem obvious. However, unlike in
a scalar field theory, the interactions in a non-Abelian gauge theory
involve time derivatives in the mixed space (for example, the three
gluon vertex or the ghost interaction vertex) and, in principle, may
complicate the general proof of the thermal operator representation of
an arbitrary graph. On the other hand, we note that the basic thermal
operator (\ref{tob}) in the factorization of the propagators
(\ref{gaugefactorization}) is independent of time coordinates. As a
result, it follows trivially that
\begin{equation}
\partial_{t} {\cal O}_{\rm B}^{(T)} (E) = {\cal O}_{\rm B}^{(T)} (E)
\partial_{t},\label{commutativity}
\end{equation}
so that the basic thermal operators in a propagator can be trivially
commuted past the derivatives in the vertices leading to a thermal
operator representation of any arbitrary graph. Let us illustrate this
with the example of the one loop graph (see Fig. \ref{f2}) depicting the
ghost contribution to the gauge self-energy.
\noindent In this case, the graph can be written as (we suppress all
multiplicative factors associated with the diagram for simplicity,
identify $k_{3}=k_{1}$ and
look only at the contributions coming from a time derivative in the
vertex which may present a challenge to a thermal operator representation)
\begin{widetext}
\begin{eqnarray}
\Gamma_{2}^{ab (T)} & = &  \int \prod_{i=1}^{2}\frac{d^{3}k_{i}}{(2\pi)^{3}}\
  (2\pi)^{3}\delta^{3} (k_{i}+p_{i}+k_{i+1}) \gamma_{2}^{ab
  (T)}\nonumber\\
& = & \int
  \prod_{i=1}^{2}\frac{d^{3}k_{i}}{(2\pi)^{3}}\ (2\pi)^{3}
\delta^{3}(k_{i}+p_{i}+k_{i+1}) f^{apc}f^{bqr}
  \partial_{t_{1}}D_{++}^{qc (T)} (t_{2}-t_{1}, E_{1})
  \partial_{t_{2}} D_{++}^{pr (T)} (t_{1}-t_{2}, E_{2})\nonumber\\
 & = &  \int \prod_{i=1}^{2} 
  \frac{d^{3}k_{i}}{(2\pi)^{3}}\ (2\pi)^{3}
  \delta^{3} (k_{i}+p_{i}+k_{i+1}) f^{apc}f^{bqr}
  \!\!\!\left(\partial_{t_{1}} {\cal O}_{\rm B}^{(T)} (E_{1}) D_{++}^{qc
  (0)} (t_{2}-t_{1}, E_{1})\right)\!\!
  \left(\partial_{t_{2}} {\cal O}_{\rm B}^{(T)} (E_{2})  D_{++}^{pr
  (0)} (t_{1}-t_{2}, E_{2})\right)\nonumber\\
 & = &  \int \prod_{i=1}^{2} 
  \frac{d^{3}k_{i}}{(2\pi)^{3}}\ (2\pi)^{3}
  \delta^{3} (k_{i}+p_{i}+k_{i+1})\ {\cal O}_{\rm
  B}^{(T)} (E_{1}) {\cal O}_{\rm B}^{(T)} (E_{2}) f^{apc}f^{bqr}
  \partial_{t_{1}}D_{++}^{qc (0)} (t_{2}-t_{1}, E_{1})
  \partial_{t_{2}} D_{++}^{pr (0)} (t_{1}-t_{2}, E_{2})\nonumber\\
 & = & \int \prod_{i=1}^{2} 
  \frac{d^{3}k_{i}}{(2\pi)^{3}}\ (2\pi)^{3}
  \delta^{3} (k_{i}+p_{i}+k_{i+1})\ {\cal O}_{\rm
  B}^{(T)} (E_{1}) {\cal O}_{\rm B}^{(T)} (E_{2})\
  \gamma_{2}^{ab (0)}\nonumber\\
& = & \int \prod_{i=1}^{2} 
  \frac{d^{3}k_{i}}{(2\pi)^{3}}\ (2\pi)^{3}
  \delta^{3} (k_{i}+p_{i}+k_{i+1})\ {\cal O}^{(T)}\ \gamma_{2}^{ab (0)},
\end{eqnarray}
\end{widetext}
where in the last step we have identified the thermal operator for the
graph to be
\begin{equation}
{\cal O}^{(T)} = {\cal O}_{\rm B}^{(T)} (E_{1}) {\cal O}_{\rm
  B}^{(T)} (E_{2}).
\end{equation}
Thus, we see that in spite of the presence of time derivatives in the
vertices in the mixed space representation, the thermal operator
representation holds simply because the basic thermal operator
commutes with the time derivative operator. With this observation, it
is clear that for an interacting
non-Abelian gauge theory, we can write the thermal operator
representation for any arbitrary $N$-point graph at finite temperature
(involving gauge and ghost vertices of ``$\pm$'' type) at any loop as
\begin{eqnarray}
& &  \Gamma_{N}^{a_{1}\cdots a_{N} (T)}
= \int \prod_{i=1}^{I}
  \frac{d^{3}k_{i}}{(2\pi)^{3}} \prod_{v=1}^{V}
  (2\pi)^{3} \delta_{v}^{(3)} (k,p) \gamma_{N}^{a_{1}\ldots a_{N}
  (T)}\nonumber\\ 
& = &\!\! \int \prod_{i=1}^{I} \frac{d^{3}k_{i}}{(2\pi)^{3}} \prod_{v=1}^{V}
  (2\pi)^{3} \delta_{v}^{(3)} (k,p) {\cal O}^{(T)}
  \gamma_{N}^{a_{1}\ldots a_{N} (0)},
\end{eqnarray}
where we have suppressed the Lorentz indices associated with the graph
and the thermal operator for the graph has the form
\begin{equation}
{\cal O}^{(T)} = \prod_{i=1}^{I} {\cal O}_{\rm B}^{(T)} (E_{i}).
\end{equation}
In a similar manner, it can be shown that in the absence of chemical
potentials any diagram in an interacting theory involving gauge fields,
fermions and scalar fields at finite temperature will have a thermal
operator representation. The interesting and challenging case,
however, seems to be in the presence of a chemical potential which we
discuss in the next section.

\section{Fermions with a chemical potential}

Let us next consider QED at finite temperature and density. In the
Feynman gauge, the theory is described by the Lagrangian density
\begin{equation}
{\cal L} = - \frac{1}{4} F_{\mu\nu} F^{\mu\nu} +
\bar{\psi}\left(i D\!\!\!\!\slash - m\right)\psi - \frac{1}{2}
\left(\partial\cdot A\right)^{2} + \mu
\bar{\psi}\gamma^{0}\psi,\label{qedlagrangian}
\end{equation}
where $\mu$ represents the chemical potential associated with the
fermion and the covariant derivative is defined to be
\begin{equation}
D_{\mu}\psi = \partial_{\mu}\psi - ie A_{\mu}\psi.
\end{equation}
In (\ref{qedlagrangian}) we have neglected the free Lagrangian density
for the ghosts which is not relevant for our discussions. 

As we have argued before, there is no chemical potential associated
with the photon. As a result, the propagator for the gauge boson in
the Feynman gauge will continue to be what we have discussed in the
last section (without any internal indices). On the other hand, in
momentum space in the closed time path formalism, the components of
the fermion propagator in the presence of a chemical potential take
the forms
\begin{widetext}
\begin{eqnarray}
S_{++}^{(T,\mu)} (p) & = & (p\!\!\!\slash + m + \mu
\gamma^{0})\Big[L(\epsilon)\,
\frac{i}{(p_{0}+\mu)^{2} - E^{2}+i\epsilon}
 - 2\pi n_{\rm F} ({\rm sgn}(p_{0}+\mu) p_{0})
  \delta ((p_{0}+\mu)^{2}-E^{2})\Big],\nonumber\\
S_{+-}^{(T,\mu)} (p) & = & 2\pi(p\!\!\!\slash + m + \mu
\gamma^{0})\left(\theta (-p_{0}-\mu) - n_{\rm F} ({\rm sgn}
(p_{0}+\mu) p_{0})\right)\delta ((p_{0}+\mu)^{2}-E^{2}),\nonumber\\
S_{-+}^{(T,\mu)} (p) & = & 2\pi(p\!\!\!\slash + m + \mu
\gamma^{0})\left(\theta (p_{0}+\mu) - n_{\rm F}
({\rm sgn} (p_{0}+\mu) p_{0})\right)\delta
((p_{0}+\mu)^{2}-E^{2}),\label{Smu}\\
S_{--}^{(T,\mu)} (p) & = & (p\!\!\!\slash + m + \mu
\gamma^{0})\Big[-L(\epsilon)\,
\frac{i}{(p_{0}+\mu)^{2} - E^{2}-i\epsilon}
 - 2\pi n_{\rm F} ({\rm sgn}(p_{0}+\mu) p_{0})
  \delta ((p_{0}+\mu)^{2}-E^{2})\Big],\nonumber
\end{eqnarray}
\end{widetext}
where $E = \sqrt{\vec{p}^{\ 2} + m^{2}}$. Equation (\ref{Smu}) clearly reduces
to (\ref{S}) when $\mu = 0$. The Fourier transform of (\ref{Smu}) in
the $p_{0}$ variable leads to the propagator in the mixed space which
can be seen to have a nontrivial factorization (as is the case with the
complex scalar field discussed in ref. {\bf I})
\begin{equation}
S_{\alpha\beta}^{(T,\mu)} (t, \vec{p}\ ) = e^{i\mu t} {\cal O}_{\rm
  F}^{(T,\mu)} (E) S_{\alpha\beta}^{(0,0)} (t,\vec{p}\
  ),\label{factorizationmu}
\end{equation}
where, as before, $\alpha,\beta = \pm$ and the basic thermal operator
in (\ref{factorizationmu}) has the form
\begin{eqnarray}
{\cal O}_{\rm F}^{(T,\mu)} (E) & = & 1 - \frac{n_{\rm F}^{+} + n_{\rm
    F}^{-}}{2} (1 - S(E))\nonumber\\
 &  & \quad + \frac{n_{\rm F}^{+} - n_{\rm F}^{-}}{2} (1 + S(E))
    \frac{i\partial_{t}}{E},\label{tofmu}
\end{eqnarray}
where we have defined
\begin{equation}
n_{\rm F}^{\pm} = n_{\rm F} (E\pm \mu),
\end{equation}
and $S_{\alpha\beta}^{(0,0)} (t,\vec{p}\ )$ represent the components
of the fermion propagator at zero temperature and zero chemical
potential given in (\ref{fermioncomponents}). 

We note that the
structure of the basic thermal operator in (\ref{tofmu}) is quite
analogous to the case of the complex scalar field with a chemical
potential discussed in ref. {\bf I}. This is a scalar operator which
involves a time derivative operator (it does not depend on the time
coordinate). Unlike in the case of the scalar field, however, in this
case, we can equivalently define a basic thermal operator which is
independent of the time derivative, but instead is a matrix (in the
Dirac space), namely, in this case we can also write
\begin{equation}
S_{\alpha\beta}^{(T,\mu)} (t, \vec{p}\ ) = e^{i\mu t} \tilde{{\cal O}}_{\rm
  F}^{(T,\mu)}  S_{\alpha\beta}^{(0,0)} (t,\vec{p}\
  ),\label{factorizationmu1}
\end{equation}
where
\begin{equation}
\tilde{\cal O}_{\rm F}^{(T,\mu)} = 1 - \left(n_{\rm F}^{-} A(E) - n_{\rm
  F}^{+} B(E)\right)\frac{\gamma^{0}}{2E} (1 - S (E)),\label{tofmu1}
\end{equation}
where $A(E), B(E)$ are matrices defined in (\ref{ab}). The two forms
of the basic thermal operator are related through the first order
equation satisfied by the fermion propagator in the mixed
space. However, the 
scalar form of the basic thermal operator (in spite of the time
derivative operator) is easier to use than the matrix one and we will
carry out our discussions in terms of the factorization
(\ref{factorizationmu}). We note that both forms of the basic thermal
operator in (\ref{tofmu}) and (\ref{tofmu1}) can be checked to be
projection operators, namely,
\begin{equation}
\left({\cal O}_{\rm F}^{(T,\mu)}\right)^{2} = {\cal O}_{\rm
  F}^{(T,\mu)},\quad \left(\tilde{{\cal O}}_{\rm F}^{(T,\mu)}\right)^{2}
  = \tilde{{\cal O}}_{\rm F}^{(T,\mu)},
\end{equation}
and enforce the necessary anti-periodicity in the present case.

As in the case of the complex scalar field with a chemical potential
(in ref. {\bf I}), here we note that the presence of the time derivative
term in the basic thermal operator (\ref{tofmu}) makes it difficult to
give a general proof of a thermal operator representation for any
arbitrary graph, particularly in cases involving internal time
coordinates which are integrated over. In the scalar theory, we had
shown in a specific complicated graph that a (nontrivial) thermal operator
representation results in spite of the presence of the time derivative
operator and this led to the hope that a thermal operator
representation may result in general. In the fermionic theory, we will
show through an explicit calculation that the thermal operator
representation fails in the presence of a chemical potential and we
will trace this failure (in the next section) to a renormalization of
the chemical potential because of radiative corrections. For the
purpose of explicit calculations, we will use the imaginary time
formalism \cite{das:book97,kapusta:book89,lebellac:book96} where there
is no doubling of fields. In this case, the
factorization of the thermal propagator (\ref{factorizationmu}) can be
written as
\begin{equation}
S^{(T,\mu)} (\tau, \vec{p}\ ) = e^{\mu\tau}\ 
{{\cal O}}_{\rm F}^{(T,\mu)} (E) S^{(0,0)} (\tau,\vec{p}\
  ),\label{imaginarytfactorization} 
\end{equation}
where ${{\cal O}}_{\rm F}^{(T,\mu)} (E)$ is the rotation of the
operator in (\ref{tofmu}) to imaginary time
\begin{eqnarray}\label{49o}
{{\cal O}}_{\rm F}^{(T,\mu)} (E) & = & 1 - \frac{n_{\rm F}^{+} +
  n_{\rm F}^{-}}{2} (1 - S(E))\nonumber\\
 &  & \qquad - \frac{n_{\rm F}^{+} - n_{\rm F}^{-}}{2} (1 + S(E))
  \frac{\partial_{\tau}}{E},
\end{eqnarray}
and the propagator at zero temperature in the absence of a chemical
potential has the form
\begin{equation}
S^{(0,0)} (\tau,\vec{p}\ ) = \frac{1}{2E}\left[\theta(\tau) A(E)
  e^{-E\tau} + \theta(-\tau) B(E) e^{E\tau}\right].
\end{equation}
Here $A(E),B(E)$ denote the Euclidean rotation of the
matrices in (\ref{ab}), namely,
\begin{equation}
A(E) = i\gamma_{0}E - \vec{\gamma}\cdot \vec{p} + m,\quad
B(E) = -i\gamma_{0}E - \vec{\gamma}\cdot \vec{p} + m.
\end{equation}

\begin{figure}[h!]
\includegraphics[scale=0.5]{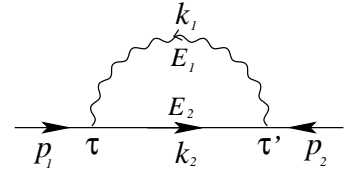}
\caption{The fermion self-energy at one-loop order.}
\label{f3}
\end{figure}

Let us next analyze the one loop fermion self-energy graph at finite
temperature 
and density in QED (see Lagrangian density (\ref{qedlagrangian})) at
one loop in the imaginary time formalism. The two point function in
Fig. \ref{f3}  can be explicitly evaluated and leads to (we note that
the fermion self-energy is simply the two point function with momentum
conserving delta functions factored out and we identify $k_{3}=k_{1}$
in the derivation below)
\begin{eqnarray}
& &\Gamma_{2}^{(T,\mu)} = \!\!\int\!\! \prod_{i=1}^{2}
\frac{d^{3}k_{i}}{(2\pi)^{3}}\ (2\pi)^{3} \delta^{3}
(k_{i}+p_{i}-k_{i+1})\ \tilde{\gamma}_{2}^{(T,\mu)}\nonumber\\
& & \quad = (2\pi)^{3}\delta^{3}(p_{1}+p_{2}) \int
\frac{d^{3}k}{(2\pi)^{3}}\ \tilde{\gamma}_{2}^{(T,\mu)},\label{selfenergy}
\end{eqnarray}
where we have identified the integrand with
$\tilde{\gamma}_{2}^{(T,\mu)}$ (to avoid confusion with the Dirac
gamma matrices) and have carried out the integration over $\vec{k}_{2}$
using one of the delta functions (and have identified
$\vec{k}_{1}=\vec{k}$). The integrand has the explicit form
\begin{eqnarray}
\tilde{\gamma}_{2}^{(T,\mu)} & = & -e^{2}\gamma_{\mu}S^{(T,\mu)}
(\tau-\tau',\vec{k}_{2}\ )\gamma_{\mu} D^{(T)} (\tau-\tau',
E_{1})\nonumber\\
& = & -e^{2} \Big[e^{\mu(\tau-\tau')} {\cal O}_{\rm B}^{(T)} (E_{1})
\bar {\cal O}_{\rm F}^{(T,\mu)} (E_{2})\nonumber\\
&  & \!\!\!\times \left(\gamma_{\mu} S^{(0,0)}
    (\tau-\tau',\vec{k}_{2}\ )\gamma_{\mu} D^{(0)} (\tau-\tau',
  E_{1})\right)\nonumber\\
&  & \!\!\! + \frac{2i(n_{\rm F}^{+}(E_{2}) - n_{\rm
      F}^{-} (E_{2}))\gamma_{0}}{E_{1}^{2} - E_{2}^{2}} \delta
  (\tau-\tau')\Big]\nonumber\\
& = &  e^{\mu(\tau-\tau')}{\cal O}_{\rm B}^{(T)} (E_{1}) 
\bar {\cal O}_{\rm F}^{(T,\mu)}(E_1,E_2)\, \tilde{\gamma}_{2}^{(0,0)}\nonumber\\
&  &\quad  - \frac{2ie^{2}(n_{\rm F}^{+}(E_{2})-n_{\rm F}^{-} 
(E_{2}))\gamma_{0}}{E_{1}^{2}-E_{2}^{2}} \delta(\tau-\tau'),\label{contact}
\end{eqnarray}
where the modified fermion operator $\bar {\cal O}_{\rm F}^{(T,\mu)}$ is similar to the one in 
(\ref{49o}) except that it involves $\partial_\tau/(E_1+E_2)$ and 
\begin{equation}
E_{1} = |\vec{k}|,\quad E_{2} = \sqrt{(\vec{k}+\vec{p}\ )^{2} +
  m^{2}},\quad \vec{p}_{1}=\vec{p},\quad \vec{k}_{2} = \vec{k} +
  \vec{p}. 
\end{equation}
Thus, we see explicitly from (\ref{contact}) that,
unlike the case of the complex scalar field discussed in ref. \cite{silvana},
here a thermal operator representation for the
fermion two point function at one loop breaks down in the presence of
a chemical potential. The additional term leading to the breakdown of
the thermal operator representation is a contact term which vanishes
for $\mu\rightarrow 0$ since
\begin{equation}
\lim_{\mu\rightarrow 0}\ (n_{\rm F}^{+} (E_{2}) - n_{\rm F}^{-}
(E_{2}))  \rightarrow 0.
\end{equation}

\section{Renormalization of Chemical Potential}

Let us note that the chemical potential can be
thought of as a constant background electrostatic potential. This is
particularly clear if we note that the free part of the fermion
Lagrangian density in (\ref{qedlagrangian}) can be written in the
Euclidean space as
\begin{equation}
{\cal L}_{f, {\rm free}} =
\bar{\psi}\left(\gamma_{0}(\partial_{\tau}-\mu) + \vec{\gamma}\cdot
\vec{\nabla}\right)\psi + m\bar{\psi}\psi.
\end{equation}
As a result of this structure of the theory, one can derive certain
identities as follows. Adding the sources for the fields, we can write
the generating functional for the theory in (\ref{qedlagrangian}) (in
imaginary time) as
(we set $\hbar = 1$)
\begin{eqnarray}
&  & Z[\mu,J_{\mu},\eta,\bar{\eta}] =
 e^{-W[\mu,J_{\mu},\eta,\bar{\eta}]}\nonumber\\
& = & \!\!
\int\!\! {\cal D}A_{\mu} {\cal D}\bar{\psi} {\cal D}\psi\
e^{-\int_{0}^{\frac{1}{T}}d\tau\! \int\! d^{3}x
  \left({\cal L} + J_{\mu}A_{\mu} + i (\bar{\eta}\psi -
  \bar{\psi}\eta)\!\right)}.
\end{eqnarray}
This, in turn, leads to the identity
\begin{eqnarray}
\frac{\partial W}{\partial \mu} & = & \int_{0}^{\frac{1}{T}}\!\!d\tau\!\! \int
  d^{3}x  \left(i\frac{\delta
  W}{\delta \eta (x)} \gamma_{0}\frac{\delta W}{\delta
  \bar{\eta}(x)}\right.\nonumber\\ 
&  & \qquad\qquad\quad\left.  - i {\rm Tr} \gamma_{0} \frac{\delta^{2}W}{\delta
  \eta (x)\delta  \bar{\eta} (x)}\right).\label{generatingidentity}
\end{eqnarray}
By taking the second order derivative with respect to
$\eta,\bar{\eta}$ and setting all sources to zero, this leads to the
identity (when we Fourier transform the spatial coordinates)
\begin{eqnarray}
&  & \frac{\partial S^{(T,\mu)}(\tau_{1}-\tau_{2},\vec{p}\ )}{\partial
 \mu}\nonumber\\
&  & = i\!
\int_{0}^{\frac{1}{T}}\!\!\!\! d\tau\ S^{(T,\mu)}(\tau_{1}-\tau,\vec{p}\
  )\gamma_{0} S^{(T,\mu)} (\tau-\tau_{2},\vec{p}\ ).\label{identity}
\end{eqnarray}
Such an identity has already proved quite useful in the solution of
the $0+1$ dimensional Chern-Simons QED \cite{Das:1997gg} and with the
explicit form of
the fermion propagator in (\ref{imaginarytfactorization}) it can be
checked that this is true. It also follows from (\ref{identity}) that 
\begin{equation}
\frac{1}{n!} \frac{\partial^{n}S^{(T,\mu)}}{\partial \mu^{n}} =
(i)^{n} S^{(T,\mu)}\gamma_{0}\cdots
S^{(T,\mu)}\gamma_{0}S^{(T,\mu)},\label{identity1}
\end{equation}
where there are $n$ insertions of $\gamma_{0}$ on the right hand side
and we have used a compact notation suppressing the internal time
integrations.

Using the above result, we can see that a correction to the chemical
potential may arise from successive insertions of the operator 
$\delta\mu\,\gamma_0$ in the fermion propagator $S^{(T,\mu)}$. The
resulting corrected propagator $S^{(T,\mu+\delta \mu)}$ is then
obtained by summing the geometric series
\begin{eqnarray}
&\sum_{n=0}^{\infty} (i)^n S^{(T,\mu)}\,\left(\delta\mu\,\gamma_0\right)
\cdots                  S^{(T,\mu)}\,\left(\delta\mu\,\gamma_0\right) \,S^{(T,\mu)}
 =  \nonumber \\
&\displaystyle{
\sum_{n=0}^{\infty} (i)^n \frac{(\delta\mu)^n}{n!}
                       \frac{\partial^n S^{(T,\mu)}}{\partial \mu^n}
=  S^{(T,\mu+\delta \mu)}}.
\end{eqnarray}
Thus, the effect of such insertions is to shift the chemical potential
to an effective value given by $\mu+\delta\mu$. In order to evaluate
this shift (finite renormalization) we have to perform a more
systematic analysis of the fermion self-energy. To this end, let us
calculate the complete self-energy in momentum space
\cite{das:book97}  (namely, Fourier
transform (\ref{selfenergy}) in the time variable and factor out overall
energy-momentum conserving delta functions)
\begin{equation}
\Sigma^{(T,\mu)} (p) = \frac{1}{2}\int_{-\frac{1}{T}}^{\frac{1}{T}} d\tau\
e^{ip_{0}\tau}\ \Sigma^{(T,\mu)} (\tau,\vec{p}\ ).
\end{equation}
A direct evaluation yields
\begin{widetext}
\begin{eqnarray}
\Sigma^{(T,\mu)} (p) & = & -\frac{1}{2}\int
  \frac{d^{3}k}{(2\pi)^{3}}
  \frac{e^{2}}{4E_{1}E_{2}}\left[\bar{A}\left(\frac{(1+n_{\rm B} 
  (E_{1})- n_{\rm
  F}^{-}(E_{2}))}{E_{1}+E_{2}-(ip_{0}+\mu)} -
  \frac{(n_{\rm B}(E_{1})+n_{\rm F}^{-} (E_{2}))}{E_{1}-E_{2} +
  (ip_{0}+\mu)}\right)\right.\nonumber\\
&  & \qquad\qquad\qquad\left. +\bar{B}\left( \frac{(1+ n_{\rm
  B}(E_{1})-n_{\rm
  F}^{+}(E_{2}))}{E_{1}+E_{2}+(ip_{0}+\mu)} -
  \frac{(n_{\rm B}(E_{1}) + n_{\rm F}^{+}(E_{2}))}{E_{1}-E_{2}-
  (ip_{0}+\mu)}\right)\right],\label{sigma} 
\end{eqnarray}
\end{widetext}
where we have identified
\begin{eqnarray}
\bar{A} & = & 2(i\gamma_{0}E_{2} - \vec{\gamma}\cdot \vec{k}_{2} -
2m)\nonumber\\
\bar{B} & = & 2(-i\gamma_{0}E_{2} - \vec{\gamma}\cdot \vec{k}_{2} -
2m).
\end{eqnarray}
It is clear from the explicit form of (\ref{sigma}) that the
self-energy (in fact, any thermodynamic function) is a function of
$(ip_{0}+\mu,\mu)$ where the extra $\mu$ dependence comes from the
explicit dependence of the fermion distribution functions on the
chemical potential.
The analytic continuation of the self-energy can be carried out
appropriately \cite{jeon} through
\begin{equation}
iP_{0} = ip_{0}+\mu\rightarrow P_{0}= (p_{0}+\mu)(1 + i\epsilon),
\end{equation} 
say for the time ordered amplitude.
We recognize from (\ref{sigma}) that the real part of the coefficient
of the 
$\gamma^{0}$ term in (\ref{sigma}) coincides
  exactly with the coefficient of the contact term in the mixed space
  in (\ref{contact}) when  $p_{0}+\mu=0$. 

In fact, let us note that after analytic continuation, we can write
the coefficient of the $\gamma^{0}$ term in (\ref{sigma}) as 
\begin{widetext}
\begin{eqnarray}
\frac{1}{4} {\rm Tr}\ \gamma^{0} \Sigma^{(T,\mu)} (p) &=&
  -\frac{e^{2}}{2}\int \frac{d^{3}k}{(2\pi)^{3}}
  \frac{1}{E_{1}}\left[\left\{\frac{1}{E_{1}+ E_{2} - P_{0}} -
  \frac{1}{E_{1}+E_{2} + P_{0}}\right.\right.\nonumber\\
& & + n_{\rm
  B}(E_{1})\left(\frac{1}{E_{1}+E_{2}-P_{0}} -
  \frac{1}{E_{1}+E_{2}+P_{0}} + \frac{1}{E_{1}-E_{2}-P_{0}} -
  \frac{1}{E_{1}-E_{2}+P_{0}}\right)\nonumber\\
&  &\left.-\frac{n_{\rm F}^{+}(E_{2}) + n_{\rm
  F}^{-}(E_{2})}{2}\left(\frac{1}{E_{1}+E_{2}-P_{0}}
-\frac{1}{E_{1}+E_{2}+P_{0}} -\frac{1}{E_{1}-E_{2}-P_{0}} +
  \frac{1}{E_{1}-E_{2}+P_{0}}\right)\right\}\nonumber\\
 &  &\left. + \frac{n_{\rm
  F}^{+}(E_{2})-n_{\rm F}^{-}(E_{2})}{2}\left(\frac{1}{E_{1}+E_{2}-P_{0}}
+\frac{1}{E_{1}+E_{2}+P_{0}} +\frac{1}{E_{1}-E_{2}-P_{0}} +
  \frac{1}{E_{1}-E_{2}+P_{0}}\right)\right].\label{expression}
\end{eqnarray}
\end{widetext}
Let us denote the terms involving the braces as $\Sigma_{1}$ and the terms in
the last parenthesis as $\Sigma_{2}$ (after integration) so that we can
write
\begin{equation}
\frac{1}{4} {\rm Tr} \gamma^{0} \Sigma^{(T,\mu)} (p) =
\Sigma_{1}(P_{0},\vec{p},\mu) + \Sigma_{2} (P_{0},\vec{p},\mu).
\end{equation}
From the explicit form of the terms in (\ref{expression}), we note that
under $P_{0}\rightarrow - P_{0},\mu\rightarrow -\mu$, both
$\Sigma_{1},\Sigma_{2}$  change
sign. However, since $\Sigma_{1}$ is manifestly symmetric under
$\mu\rightarrow -\mu$, it is anti-symmetric under $P_{0}\rightarrow
-P_{0}$ and, consequently, vanishes when $P_{0}=0$. On the other hand,
$\Sigma_{2}$ is anti-symmetric under $\mu\rightarrow -\mu$ and,
consequently, is symmetric under $P_{0}\rightarrow -P_{0}$ and does not
vanish when $P_{0}=0$ and, in fact, yields the contact term in
(\ref{contact}). Using the results in appendix {\bf B}, we can
evaluate  $\Sigma_{2}$ explicitly in
the high $T$ limit and when $\vec{p}=0$, it has the form (for $T\gg m$)
\begin{equation}\label{68x}
\Sigma_{2} (P_{0}=0,\vec{p}=0) \simeq -\frac{e^{2}\mu (\pi^{2} T^{2} +
  \mu^{2})}{6\pi^{2} m^{2}}.
\end{equation}
This has a quadratic mass singularity that arises only when
$\mu\neq 0$ (and, therefore, at finite charge density). 
The self-energy at zero temperature and chemical potential
on the other hand, is infrared finite \cite{Zuber:book}
(logarithmic infrared divergences only appear if one expands around the
singular point $p^2=m^2$). Consequently, the singular 
term (\ref{68x}) cannot
be related to the zero temperature fermion self-energy through a
regular  thermal operator. We believe that the presence of this strongly divergent
infrared behavior is responsible for the failure of the thermal
operator representation in the case of a nonvanishing chemical potential.

To understand the renormalization of the chemical potential, we have
to analyze the poles of the fermion propagator. We note that with the
self-energy corrections, the complete two point function at one loop
(in Minkowski space) can be written as
\begin{equation}
i(S^{ (T,\mu)})^{-1} (p) = p\!\!\!\slash -m + \mu \gamma^{0} - \Sigma
(p,T,\mu),\label{pole}
\end{equation}
where $\Sigma$ can only be calculated in some limit such as the high
$T$ limit. The analysis of the poles of the propagator in this limit, even
in the absence of a chemical potential, is highly nontrivial. It is
known in the absence of a chemical potential \cite{lebellac:book96}
that in the leading order at high temperature ($T\gg m$), the fermion
propagator has an absolute pole at $\vec{p}=0, p_{0} = \pm m_{f}$ where
$m_{f} = \frac{eT}{2\sqrt{2}}$ represents the thermal mass of the
fermion. For nonzero $\vec{p}$, the fermion propagator has only
partial poles corresponding to two quasi-particle modes. Here we will
follow the same analysis restricting ourselves to only the absolute
pole in the leading order at high temperature. 
Also, we will disregard those thermal corrections which yield a
finite renormalization of the vacuum fermion mass, because such
non-leading terms are not relevant for the analysis of the
renormalization of the chemical potential.
In the leading order at
high temperature ($T\gg m$), the terms in (\ref{sigma}) which are even
under $\mu\rightarrow -\mu$ lead to
\begin{eqnarray}
\Sigma_{\rm even} & = & \frac{m_{f}^{2}\gamma^{0}}{2p} \ln
\frac{P_{0}+p}{P_{0}-p}\nonumber\\
 &  & + \frac{m_{f}^{2} \vec{\gamma}\cdot\hat{p}}{p}\left(1 -
\frac{P_{0}}{2p} \ln
\frac{P_{0}+p}{P_{0}-p}\right),\label{nonanalytic}
\end{eqnarray}
where we have defined $p=|\vec{p}|$ and $\hat{p}$ denotes the unit
vector along $\vec{p}$. In the presence of a chemical potential we
have
\begin{equation}
m_{f}^{2} = \frac{e^{2}}{8}\left(T^{2} +
 \frac{\mu^{2}}{\pi^{2}}\right),\label{mf}
\end{equation} 
which is well known (see, for example, \cite{lebellac:book96}).
The non-analytic behavior of $\Sigma_{\rm even}$ at $P_{0}=0, p=0$ is
obvious from (\ref{nonanalytic}). We also note that all the
$\vec{\gamma}\cdot \hat{p}$ terms in (\ref{nonanalytic}) vanish in the
  limit $p=0$ (basically because in this case, there is no direction
  available to contract the gamma matrix) so that we can write
\begin{equation}
\Sigma_{\rm even} (\vec{p}=0) \simeq
\frac{m_{f}^{2}}{P_{0}} \gamma^{0} = \gamma^{0} \Sigma_{1}
(P_{0},\vec{p}=0). \label{sigmaeven}
\end{equation} 
The absolute pole in the
fermion propagator continues to be at $\vec{p}=0$ and at this point,
the terms proportional to $\gamma_0$
in (\ref{sigma}) which are odd under $\mu\rightarrow -\mu$
yield 
\begin{eqnarray}
\Sigma_{\rm odd} & = & -
\frac{e^{2}(P_{0}^{2}+m^{2})\gamma^{0}}{2\pi^{2}}\!\! 
\int_{0}^{\infty} \! dk k^{2}\ \frac{(n_{\rm F}^{-}(k)-n_{\rm
    F}^{+}(k))}{4k^{2}P_{0}^{2} -
  (P_{0}^{2}-m^{2})^{2}}\nonumber\\
 & = & \gamma^{0} \Sigma_{2} (P_{0},\vec{p}=0).
\end{eqnarray}
Using the results in appendix {\bf B}, this can be evaluated in the
high temperature limit and shows that at
$|P_{0}| = |M|\sim e\, T \gg m$ it is well behaved and has the value
\begin{eqnarray}
\Sigma_{\rm odd} (P_{0}=M,\vec p =0) & = & \gamma^{0}\Sigma_{2}(P_{0}=M,
\vec{p}=0)\nonumber\\
& \simeq &  
\frac{e^{2}\mu\gamma^{0}}{8\pi^{2}}.\label{sigmaodd}
\end{eqnarray}

With these results, the analysis of the pole when $\vec{p}=0$ becomes
quite straightforward. We note from (\ref{pole}) that when
$\vec{p}=0$,  the propagator will have a pole provided
\begin{equation}
\gamma^{0}P_{0} - m -\Sigma_{\rm even} (\vec{p}=0) - \Sigma_{\rm
  odd}(\vec{p}=0) = 0.\label{equation}
\end{equation}
We note from (\ref{mf}) and (\ref{sigmaeven}) that at very high temperature
$m_{f}\gg m$  so that the fermion mass may be neglected in the above
equation. 
All the other
terms are proportional to $\gamma^{0}$. If we expand 
$\Sigma_{\rm odd}$ around $P_{0}=M$, the equation (\ref{equation}) takes the
form
\begin{equation}
\gamma^{0}\left(P_{0} - \frac{m_{f}^{2}}{P_{0}} -
\frac{e^{2}\mu}{8\pi^{2}}\right) - (P_{0}-M) \Sigma'_{\rm odd}
(P_{0}=M) + \cdots = 0.
\end{equation}
Here we have used (\ref{sigmaodd}) and $\Sigma'_{\rm odd}$ denotes the
derivative of $\Sigma_{\rm odd}$ with respect to $P_{0}$.
The root of this equation and, therefore, the location of the pole is
given by 
\begin{equation}
P_{0} = M = \frac{e^{2}\mu}{16\pi^{2}} \pm m_{f}\left(1 +
\frac{e^{2}\mu^{2}}{32\pi^{2}(\pi^{2} T^{2} +
  \mu^{2})}\right)^{\frac{1}{2}},
\end{equation}
which can be equivalently written as
\begin{equation}
P_{0}-\frac{e^{2}\mu}{16\pi^{2}} = \pm m_{f}\left(1 +
\frac{e^{2}\mu^{2}}{32\pi^{2}(\pi^{2} T^{2} +
  \mu^{2})}\right)^{\frac{1}{2}}.
\end{equation}
To the order that we are working, this can be simplified to give the
location of the pole at
\begin{eqnarray}
& p_{0} +  \mu\left(1 - \frac{e^{2}}{16\pi^{2}}\right) = \pm m_{f}\nonumber\\
{\rm or,} &  \\
& p_{0} + \mu_{\rm R} =  \pm \frac{e}{2\sqrt{2}}\left(T^{2} +
\frac{\mu^{2}}{\pi^{2}}\right)^{\frac{1}{2}}.\nonumber 
\end{eqnarray}
Here we have identified
\begin{equation}
\mu_{\rm R} = \mu \left(1 -
\frac{e^{2}}{16\pi^{2}}\right),\label{renormalization}
\end{equation}
which can be interpreted as the renormalized chemical potential due to
the radiative corrections of the theory. Since it is associated with a
physical pole of the propagator, we expect this result to be gauge
independent which we have explicitly checked. Such a finite renormalization
of the chemical potential has the effect of screening the
chemical potential because of thermal interactions at a finite
density. This is consistent with our earlier observation that the
chemical potential can be thought of as a constant Abelian
electrostatic potential and Abelian gauge fields lead to a screening
effect. 

\section{Summary}

In this paper, we have extended our analysis of the thermal operator
representation for Feynman graphs at finite temperature to theories
involving fermions as well gauge fields. We have shown that as long as
there is no chemical potential, the thermal operator representation
holds. We have also discussed in an appendix how a thermal operator
representation naturally arises in $0+1$ dimensional Chern-Simons QED.
However, in QED at finite temperature and density (nonzero
chemical potential), we have shown that such a factorization is
violated because of the appearance of singular contact terms.
This is explicitly worked out in the case of the
fermion self-energy at one loop. The reason for this failure of the
thermal operator representation is traced to the presence of a
quadratically divergent thermal infrared singularity in the self-energy
for a non-zero chemical potential (finite density). In this case, we
find that the chemical potential undergoes a finite renormalization
due to radiative corrections. The
renormalized chemical potential is determined from an analysis of the
pole of the fermion propagator at high temperature and shows that the
radiative corrections lead to a screening of the chemical
potential. This is argued to be consistent with the observation that a
chemical potential can be thought of as a constant electrostatic
potential and screening is a phenomenon associated with Abelian gauge
fields. 

In conclusion, we would like to point out that the lack of a complete
factorization in the presence of a chemical potential may be related to
our choice of generalizing the basic thermal operator in terms of the
simple reflection operator $S(E)$. Finding an alternate basic thermal
operator possibly dependent on other nontrivial operators and determining
its consequences on factorization is an interesting issue which is
presently under study.

\vskip .7cm

\noindent{\bf Acknowledgment}
\medskip

This work was supported in part by the US DOE Grant number DE-FG 02-91ER40685,
by MCT/CNPq as well as by FAPESP, Brazil and by CONICYT, Chile under grant
Fondecyt 1030363 and 7040057 (Int. Coop.).

\appendix

\section{0+1 Dimensional Chern-Simons QED}

As we have shown in ref. {\bf I} as well as in this paper, a thermal
operator representation for any finite temperature graph holds in any
theory in the absence of chemical potential. The reflection operator
$S(E)$ in the thermal operator simply changes $E\rightarrow -E$ and in
this way incorporates the negative energy contributions into the
graph. Furthermore, the thermal operator is independent of the time
coordinate which plays a significant role in the general proof of the
thermal operator representation. All of this is true in higher
dimensional field theories. However, in a 0+1 dimensional field theory
(quantum mechanics), the situation is different because the energy is
positive and the question arises as to whether a thermal operator
representation holds for such a theory as well. Furthermore, as is
well known, in Chern-Simons QED in 0+1 dimensions amplitudes beyond
the one point function vanish at zero temperature while all the higher
point functions are nonzero at finite temperature
\cite{Das:1997gg}. Therefore, it is
interesting to analyze how the nonzero finite temperature amplitudes
in such a theory can arise from a thermal operator acting on trivial
amplitudes. 

Let us recall that the Lagrangian for the 0+1 dimensional Chern-Simons
QED in the Euclidean space is given by
\begin{equation}
{\cal L} = \bar{\psi}(\partial_{\tau} -i A + m)\psi - i\kappa A,
\end{equation}
where $\kappa$ represents the Chern-Simons coefficient and we have set
the coupling for the gauge field to unity for simplicity. In this case,
the fermion propagator (in the imaginary time formalism) has the form
\begin{eqnarray}
S^{(T,m)} (\tau) & = & e^{-m\tau} \left(\theta(\tau) - n_{\rm F}
(m)\right)\nonumber\\
 & = & e^{-m\tau} {\cal O}_{\rm F}^{(T,m)} (\tau) S^{(0,0)} (\tau),
\end{eqnarray}
where we have identified the basic thermal operator of the theory with
\begin{equation}
{\cal O}_{\rm F}^{(T,m)} (\tau) = 1 - n_{\rm F}(m) (1 +
S(\tau)).\label{tof0+1}
\end{equation}
There are two things to note here. First, in the 0+1 dimensional case,
the mass term corresponds to a chemical potential and second, the
basic thermal operator contains a reflection operator that reflects
the time coordinate and, therefore, is manifestly time dependent. This
is quite different from the higher dimensional cases we have studied
where the basic thermal operator is independent of the time
coordinate. In the 0+1 dimensional theory, on the other hand, we do
not have higher loop diagrams (photon is non dynamical) and,
consequently, the time dependence of the basic thermal operator does
not pose a problem in deriving a thermal operator representation for
any graph. 
\begin{figure}[h!]
\includegraphics[scale=0.4]{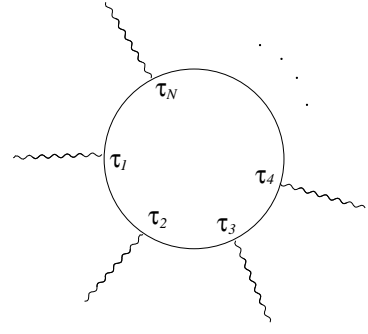}
\caption{A typical $N$-point graph in $0+1$ dimensional CS QED.}
\label{f7}
\end{figure}

For any graph of the form in Fig. \ref{f7}, we can immediately
write down the thermal operator representation as (using the
identification $\tau_{N+1}=\tau_{1}$ and the fact that the exponential
term around a closed loop vanishes)
\begin{eqnarray}
\Gamma_{N}^{(T,m)} & = & - \frac{(i)^{N}}{N!} \prod_{i=1}^{N} S^{(T,m)}
(\tau_{i}-\tau_{i+1})\nonumber\\
 & = & \prod_{i=1}^{N} {\cal O}_{\rm
  F}^{(T,m)}(\tau_{i}-\tau_{i+1})\nonumber\\
 &  & \times
\left(-\frac{(i)^{N}}{N!} \prod_{i=1}^{N} S^{(0,0)}
(\tau_{i}-\tau_{i+1})\right)\nonumber\\
& = & {\cal O}_{\rm F}^{(T,m)} \Gamma_{N}^{(0,0)},\label{0+1}
\end{eqnarray}
where we have identified
\begin{equation}
{\cal O}_{\rm F}^{(T,m)} = \prod_{i=1}^{N} {\cal O}_{\rm F}^{(T,m)}
(\tau_{i}-\tau_{i+1}).
\end{equation}
Indeed we see that formally there is a thermal operator representation
for any graph in the 0+1 dimensional Chern-Simons QED.

In practice, on the other hand, we know that in this theory
\begin{equation}
\Gamma_{N}^{(0,0)} = 0,\quad {\rm for}\ N\geq 2.
\end{equation}
The way (\ref{0+1}) works in practice is as follows. Let us identify
\begin{equation}
\tau_{i,i+1} = \tau_{i}-\tau_{i+1},
\end{equation}
and define the amplitude in (\ref{0+1}) in the limiting manner
\begin{equation}
\Gamma_{N}^{(T,m)} = \lim_{\tau_{N+1}\rightarrow \tau_{1}} {\cal
  O}_{\rm F}^{(T,m)} \Gamma_{N}^{(0,0)}.
\end{equation}
The limit $\tau_{N+1}\rightarrow \tau_{1}$ is assumed to be taken only
at the end of the calculation (after the action of the thermal
operator). With this, let us show explicitly how the correct
finite temperature two point function arises from a trivial zero
temperature amplitude.

We note that the two point amplitude at zero temperature is given by
\begin{equation}
\Gamma_{2}^{(0,0)} = -\frac{(i)^{2}}{2!} S^{(0,0)} (t_{1,2})S^{(0,0)}
(t_{2,3}) = \frac{1}{2} \theta (\tau_{12}) \theta (\tau_{2,3}).\label{zeroT}
\end{equation}
This, of course, vanishes if we identify $\tau_{3}=\tau_{1}$ (or
$\tau_{2,3}=\tau_{2,1}$). However, we are not supposed to take the
limit until the thermal operator has acted on the zero temperature
amplitude. Letting the thermal operator act on (\ref{zeroT}), we
obtain
\begin{eqnarray}
\Gamma_{2}^{(T,m)} & = & \lim_{\tau_{3}\rightarrow \tau_{1}} {\cal
  O}_{\rm F}^{(T,m)} (\tau_{12}) {\cal O}_{\rm F}^{(T,m)} (\tau_{23})
  \frac{1}{2}\theta (\tau_{12})\theta (\tau_{23})\nonumber\\
 & = & \lim_{\tau_{3}\rightarrow \tau_{1}}
  \frac{1}{2}\left(\theta(\tau_{12})-n_{\rm F} (m)\right)\left(\theta
  (\tau_{23}) - n_{\rm F} (m)\right)\nonumber\\
 & = & -\frac{1}{2} n_{\rm F} (m) \left(1-n_{\rm F} (m)\right).
\end{eqnarray}
This is indeed the correct finite temperature result (for a single
fermion flavor) \cite{Das:1997gg} and this shows
how the thermal operator correctly reproduces the nonzero finite
temperature amplitudes from trivial zero temperature ones if the
operation is carried out in a limiting manner.

\section{Derivation of the High $T$ Limit}

In this appendix, we evaluate some integrals which are used in the
text. Let us first consider
\begin{equation}
\Omega_{\rm odd}^{(0)} (\mu,m,T) = \int_{0}^{\infty} dk \left(n_{\rm
  F} (E-\mu) - n_{\rm F} (E+\mu)\right),\label{omega0}
\end{equation}
which is essential for obtaining the high
temperature limit in
(\ref{sigmaodd}). Here $E=\sqrt{\vec{k}^{\ 2} + m^{2}}$ and in the high
temperature limit $T\gg m$, it is possible to obtain a closed form
expression for $\Omega_{\rm odd}^{(0)}$ for an arbitrary $\mu$ in the
following 
way. Let us expand the integrand in a power series in $\mu$ and
integrate term by term. Every term in this series is well behaved and
leads to
\begin{equation}
\Omega_{\rm odd}^{(0)} (\mu, \frac{m}{T}\rightarrow 0) = 2\mu
\sum_{\ell=0}^{\infty} 
\frac{\mu^{2\ell}}{(2\ell+1)!} \left.\frac{\partial^{2\ell}n_{\rm F}
  (t)}{\partial t^{2\ell}}\right|_{t=0} + O \left(\frac{m}{T}\right).
\label{omega}
\end{equation}
Since the distribution function for the fermion can be expanded as
\cite{gradshteyn} 
\begin{equation}
n_{\rm F} (t) = \frac{1}{e^{t} + 1} = \frac{1}{2} \sum_{n=0}^{\infty}
E_{n} (0)\ \frac{t^{n}}{n!},
\end{equation}
where $E(x)$ represents the Euler polynomials, the expression
(\ref{omega}) can be evaluated in terms of the Euler functions. A
further simplification results from the fact that 
\begin{equation}
E_{0} (0) = 1,\quad E_{2\ell} (0) = 0\; {\rm for}\ \ell
>0. 
\end{equation}
Consequently, in this limit, (\ref{omega}) has the form
\begin{equation}
\Omega_{\rm odd}^{(0)} (\mu, \frac{m}{T}\rightarrow 0) = \mu + O
\left(\frac{m}{T}\right).\label{omega1}
\end{equation}
This result is true for any value of $\mu$.

However, when
$\frac{\mu}{T}$ is also small, one can determine the next order
correction to $\Omega^{(0)}_{\rm odd}$ as follows. We recall an alternative expansion
of the fermion distribution function as
\begin{equation}
\frac{1}{e^{z} + 1} = \frac{1}{2} - 2 \sum_{n=0}^{\infty}
\frac{z}{(2n+1)^{2}\pi^{2} + z^{2}}.
\end{equation}
If we substitute this expansion into (\ref{omega0}) and regularize the
integral by multiplying $p^{-\epsilon}$ with $\epsilon\rightarrow 0$
taken at the end \cite{dolan,haber}, the integrand can be expanded in
a power series in
$\frac{m}{T},\frac{\mu}{T}$. In this case, each term in the series can
be integrated and the series can be summed to give  Riemann's zeta
function $\zeta (2n+1,\frac{1}{2})$ \cite{gradshteyn}. A straight forward
calculation leads to
\begin{eqnarray}
& & \Omega_{\rm odd}^{(0)} (\frac{\mu}{T}\ll 1,\frac{m}{T}\ll 1) =
  \mu\nonumber\\ 
& & \quad + \mu
\sum_{n=1}^{\infty} (-1)^{n} \zeta (2n+1,\frac{1}{2}) H_{n-1}
(r^{2})\left(\frac{m}{2\pi T}\right)^{2n},\nonumber\\
& &
\end{eqnarray}
where $r^{2}=\frac{\mu^{2}}{m^{2}}$ and $H_{n} (r^{2})$ are
polynomials of order $n$ in $r^{2}$. For example, for the first few we
have
\begin{equation}
H_{0} (r^{2}) = 1,\quad H_{1} (r^{2}) = \frac{1}{2}(3 + 4r^{2}),\ldots
.
\end{equation}
In the limit, $\frac{m}{T}\rightarrow 0$, we recover (\ref{omega1}) 

Using similar techniques, we can furthermore show that
\begin{eqnarray}\label{B9x}
\Omega_{\rm odd}^{(2)} (\mu, m, T) & = & \int_{0}^{\infty} dk
k^{2}\left(n_{\rm F} (E-\mu) - n_{\rm F} (E+\mu)\right)\nonumber\\
 & = & \frac{\mu}{3}\left(\pi^{2} T^{2} +
\mu^{2}\right) + O(\frac{m}{T}),
\end{eqnarray}
\begin{eqnarray}\label{B10x}
\Omega_{\rm even}^{(1)} (\mu,m,T) & = & \int_{0}^{\infty} dk
k\left(n_{\rm F} (E-\mu) + n_{\rm F} (E+\mu)\right)\nonumber\\
 & = & \frac{1}{2}\left(\frac{\pi^{2}T^{2}}{3} +
\mu^{2}\right) + O(\frac{m}{T}).
\end{eqnarray}

One may extend the set of formulae (\ref{omega1}), (\ref{B9x}) and
(\ref{B10x}) with the help of the basic integral:
\\
\begin{equation}
\label{ipr-1}
I(p,u) = \int_0^\infty  {\frac{{x^p }}
{{e^{x - u}  + 1}}{\rm d}x}  =  - \Gamma (p + 1)Li_{p + 1} ( - e^u ),
\end{equation}
\\
which can be obtained by expanding the integrand
in powers of $e^u$ and then integrating term by term.
$Li_n(z)$ is the polylogarithm function, which is the
analytic continuation to the whole complex $z$ plane of the series
(valid for $n\ge 1$ and $|{z}|<1$)
\[
Li_n (z) = \sum\limits_{k = 1}^\infty  {\frac{{z^k }}{{k^n }}}.
\]
Using a generalization of the method employed by Haber and Weldon
in appendix A of reference \cite{haber}, one can find a power series
expansion of the function $Li_{p + 1} ( - e^u )$, leading to
\\
\begin{equation}
\label{ipr-2}
I(p,u) = \Gamma (p + 1)\sum\limits_{n = 0}^\infty
{\frac{{\left( {1 - 2^{n - p} } \right)\zeta (p + 1 - n)}}{{n!}}}\,u^n,
\end{equation}
\\
where $\zeta(z)$ is the Riemann zeta function, and where the
singular numerator corresponding to $n=p$ must be interpreted as
its limiting value, $\ln 2$. The formula is valid for all real
values of $p > -1$. Notice that for integer $p$ the series in
\eqref{ipr-2}, starting from the power $u^{p+2}$, contains powers
of the same parity only, because the zeta function vanishes at all
negative even integers.

\end{document}